\documentclass[aps,prb,twocolumn,floatfix,showpacs,citeautoscript,longbibliography,hyperlinks]{revtex4-2}
\usepackage{amsmath}
\usepackage{epstopdf}
\usepackage{amsmath}
\usepackage{amssymb}
\usepackage{amsfonts}
\usepackage[colorlinks=true,citecolor=blue]{hyperref}
\usepackage{graphicx,graphics}

\usepackage{graphicx}
\usepackage{bm}
\usepackage{hyperref}
\usepackage{braket,bbold,color}
\usepackage{tcolorbox}
\usepackage{ifthen}
\usepackage{dsfont}
\usepackage{tikz}
\usepackage{MnSymbol}
\usepackage{graphicx}
\usepackage{placeins}

\newcommand{\Ham}{\mathcal{H}}

\newcommand{\revision}[1]{{\color{black!70!black}{#1}}}

\begin{document}

\title{Lee-Yang theory of quantum phase transitions with neural network quantum states}

\author{Pascal M. Vecsei}
\affiliation{Department of Applied Physics, Aalto University, 00076 Aalto, Finland}

\author{Christian Flindt}
\affiliation{Department of Applied Physics, Aalto University, 00076 Aalto, Finland}

\author{Jose L. Lado}
\affiliation{Department of Applied Physics, Aalto University, 00076 Aalto, Finland}

\date{\today}

\begin{abstract}
Predicting the phase diagram of interacting quantum many-body systems is a central problem in condensed matter physics and related fields. A variety of quantum many-body systems, ranging from unconventional superconductors to spin liquids, exhibit complex competing phases whose theoretical description has been the focus of intense efforts. Here, we show that neural network quantum states can be combined with a Lee-Yang theory of quantum phase transitions to predict the critical points of strongly-correlated spin lattices. Specifically, we implement our approach for quantum phase transitions in the transverse-field Ising model on different lattice geometries in one, two, and three dimensions. We show that the Lee-Yang theory combined with
neural network quantum states yields predictions of the critical field, which are consistent with large-scale
quantum many-body methods. As such, our results provide a starting point for determining the phase diagram of more complex quantum many-body systems, including frustrated Heisenberg and Hubbard models.
\end{abstract}

\maketitle

\section{Introduction}

Solving a generic family of quantum many-body problems and ultimately predicting their phase diagram is a challenging task~\cite{Vojta2003,Sachdev2011}. The exponential growth of the Hilbert space with the system size, especially
for high dimensional systems, makes most realistic models intractable in practice. Some problems, such as the transverse-field Ising model in one dimension, can be solved analytically~\cite{Pfeuty1970}. However, more generally, obtaining the phase diagram of an interacting quantum many-body system is a critical open problem. To this end, several numerical tools have been developed, including Monte Carlo simulations~\cite{foulkes2001quantum}, and tensor-network algorithms~\cite{Schollwck2011thedensity}. Nevertheless, despite considerable progress, the phase diagram of many  quantum systems in two and three dimensions remain unknown~\cite{Arovas2022,Qin2022}.

\begin{figure}[b!]
 \centering
 \vspace{-25pt}
 \includegraphics[width = 0.48\textwidth]{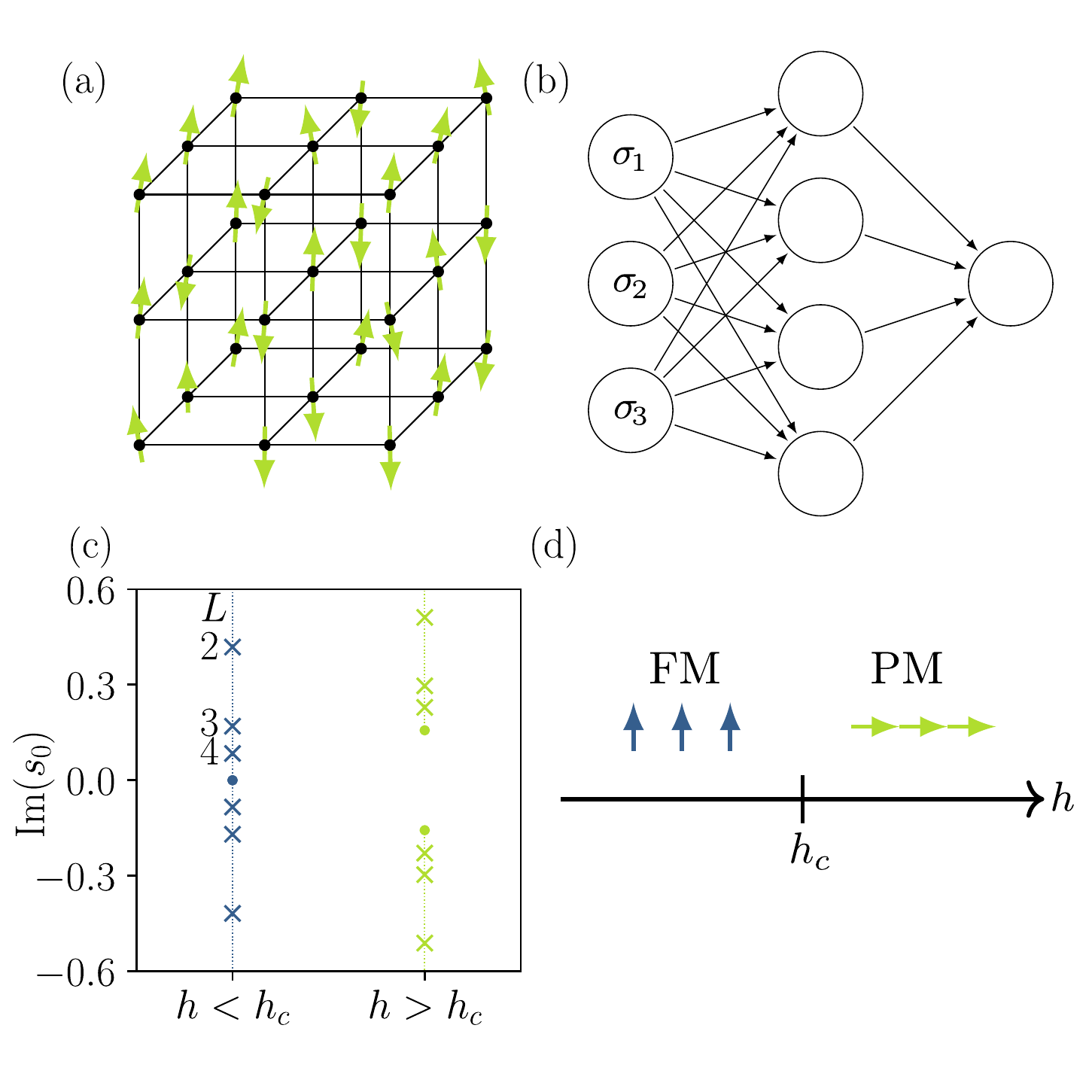}
 \vspace{-25pt}
 \caption{\label{fig:intro}  Neural network approach to quantum phase transitions. (a) Cubic Ising lattice of interacting spins in a transverse magnetic field, here a system of size $3\times 3 \times 3$. (b)  A neural network takes a configuration of the spins, encoded in the vector $\vec \sigma = (\sigma_1, ..., \sigma_N)$, and outputs the corresponding value of the wave function, $\psi_{\vec \theta}(\vec \sigma) = \langle\vec \sigma|\psi\rangle$, which depends on the variational parameters in $\vec \theta$. (c) From the fluctuations of the magnetization, we extract the zeros of the moment generating function of the magnetization \revision{(shown as $\times$)} and investigate their motion in the complex plane as we increase the system size \revision{(indicated with numbers)}. Above the critical field, $h>h_c$, the zeros remain complex in the thermodynamic limit, \revision{signaling that} the system is in the paramagnetic phase (PM). For $h<h_c$, the system is in the ferromagnetic phase (FM) with finite magnetization\revision{ and the zeros approach the origin in the thermodynamic limit. (d) Phase diagram of the transverse-field Ising model.}}
\end{figure}

Neural network quantum states
are a recently developed class of variational states~\cite{Carleo2017} that have shown
great potential for parametrizing and finding the ground state of interacting quantum many-body
systems~\cite{sharir2020deep, zen2020badcritical, wu2021unbiased, sharir2021neural, zhang2022ground, hibatallah2020recurrent,roth2021group, choo2019twodimensional, Nomura2021helping, Westerhout2020generalization, park2022expressive, szabo2020neural, roth2022high,Choo2020fermionic,Barrett2022autoregressive,adams2021variational, rigo2022solving, lovato2022hidden}. Neural network quantum states represent the wave function of a quantum many-body system as a neural network. Specifically, the neural network is a parametrized function that takes the configuration of a many-body system as the input and outputs the corresponding amplitude and phase of the wave function. By optimizing the parameters of the neural network, so that the energy is minimized, an accurate approximation of the ground state can be found. Neural network quantum states exploit the fact that neural networks can faithfully represent many complex functions~\cite{Hornik1989multilayer}, including a variety of quantum many-body wave functions. 
They have already been applied to find the wave functions of several spin models~\cite{sharir2020deep, zen2020badcritical, wu2021unbiased, sharir2021neural, zhang2022ground, hibatallah2020recurrent,PhysRevX.11.041021}, including the $J_1-J_2$ Heisenberg model~\cite{roth2021group, choo2019twodimensional, Nomura2021helping, Westerhout2020generalization, park2022expressive, szabo2020neural, roth2022high, NEURIPS2022_3173c427}. \revision{Unlike many other Monte Carlo methods, neural network quantum states, as a variant of variational Monte Carlo, can be applied to frustrated spin models.} Moreover, their use has been extended to fermionic~\cite{stokes2020phases, Choo2020fermionic} and bosonic~\cite{Saito2018machine, han2021neural, choo2018symmetries} systems, as well as to molecules~\cite{Choo2020fermionic,Barrett2022autoregressive} and nuclei~\cite{adams2021variational, rigo2022solving, lovato2022hidden}. 

In the context of critical behavior, a rigorous foundation of phase transitions was established by Lee and Yang, who considered the zeros of the partition function in the complex plane of the control parameters, for example an external magnetic field or the inverse temperature~\cite{leeYang1952StatisticalI,leeYang1952StatisticalII,blythe2003theleeyang,bena2005statistical}. This approach relies on the fact that for systems of finite size, the partition function zeros are all complex. However, if a system exhibits a phase transition, the zeros will approach the critical value on the real axis in the thermodynamic limit of large system sizes, giving rise to a non-analytic behavior of the free energy density~\cite{kim1998fisher,mulken2001classification,yamamoto2009dense,deger2018leeYang,deger2019determination, fodor2019trying,wakayama2019lee,liu2019fisher,deger2020leeYang,deger2020leeYang2,dimopoulos2022contribution,Matsumoto2022}.
Lee-Yang zeros are not just a theoretical concept, but they can also be determined experimentally~\cite{Binek1998,Wei2012,Peng2015,Flindt2013,Brandner2017}. In recent years, applications of Lee-Yang theory have been expanded to dynamical quantum phase transitions in quantum many-body systems after a quench~\cite{heyl2013dynamical,peotta2020determination,Brange2022} and to quantum phase transitions in systems at zero temperature~\cite{kist2021leeyang,vecsei2022leeyang}.

Here, we combine neural network quantum states with a Lee-Yang theory of quantum phase transitions to predict the critical behavior of interacting spin lattices in one, two, and three dimensions. As illustrated in Fig.~\ref{fig:intro}(a), we consider the transverse-field Ising model in 
different dimensions and lattice geometries. We then find the ground state of the system  as well as the fluctuations of the magnetization using neural network quantum states, Fig.~\ref{fig:intro}(b). From these fluctuations, we determine the complex zeros of the moment generating function of the magnetization and follow their motion as the system size is increased. As illustrated in Fig.~\ref{fig:intro}(c), the zeros remain complex in the thermodynamic limit in case there is no phase transition. On the other hand, if the magnetic field is tuned to its critical value, the zeros of the moment generating function will reach the real axis, signaling a phase transition. Thus, by investigating the positions of the zeros for different magnetic fields, we can map out the phase diagram of the system, Fig.~\ref{fig:intro}(d).

Our manuscript is organized as follows: In Sec.~\ref{sec:methods}, we describe the methods that we use throughout this work. In particular, we introduce the transverse-field Ising model, we discuss our calculations of the magnetization cumulants in the ground state using neural network quantum states, and we provide the details of the Lee-Yang theory that we use to predict the critical magnetic field for a given lattice geometry. In Sec.~\ref{sec:results}, we present the results of our calculations. As examples, we first discuss our procedure for the transverse-field Ising model on a one-dimensional chain, a two-dimensional square lattice, and a cubic lattice in three dimensions. We then provide predictions of the critical fields for several other lattice geometries. In Sec.~\ref{sec:discussion}, we discuss our results and the role of the coordination number and dimensionality of a given lattice. We also compare our predictions with mean-field theory, which becomes increasingly accurate in higher dimensions. Finally, in Sec.~\ref{sec:conclusion}, we summarize our conclusions. Technical details of our neural network calculations are provided in Appendix~\ref{app:details}.

\section{Methods}
\label{sec:methods}

\subsection{Transverse-field Ising model}

We consider the transverse-field Ising model on a lattice of spin-$1/2$ sites as described by the Hamiltonian 
\begin{equation}
 \hat \Ham = - J \sum_{\langle i,j \rangle} \hat \sigma_i^z \hat \sigma_j^z - h \sum_i \hat \sigma_i^x.
\end{equation}
Here, the first sum runs over all nearest neighbors, denoted by $\langle i,j \rangle$, the coupling between them is $J$, and $h$ is the transverse magnetic field. 
The one-dimensional version of this model can be solved analytically and it is known to exhibit a continuous phase transition \revision{in the thermodynamic limit} at the critical field $h_c=J$~\cite{Pfeuty1970}. Above the critical field, the system is in a paramagnetic phase with vanishing magnetization. Below it, the system exhibits spontaneous symmetry-breaking and enters a ferromagnetic phase with a non-vanishing magnetization. In the following we will investigate the model in different dimensions and geometries. 
The two-dimensional systems we consider are square, honeycomb, Kagome, and triangular lattices. In three dimensions, we consider cubic, face-centred cubic, body-centred cubic, and diamond lattices. In all of these cases, we impose periodic boundary conditions, and we compare our predictions with earlier results based on  large-scale quantum Monte Carlo simulations~\cite{bloete2002cluster}.

\begin{figure*}
 \centering
 \includegraphics[width = 1.0\textwidth]{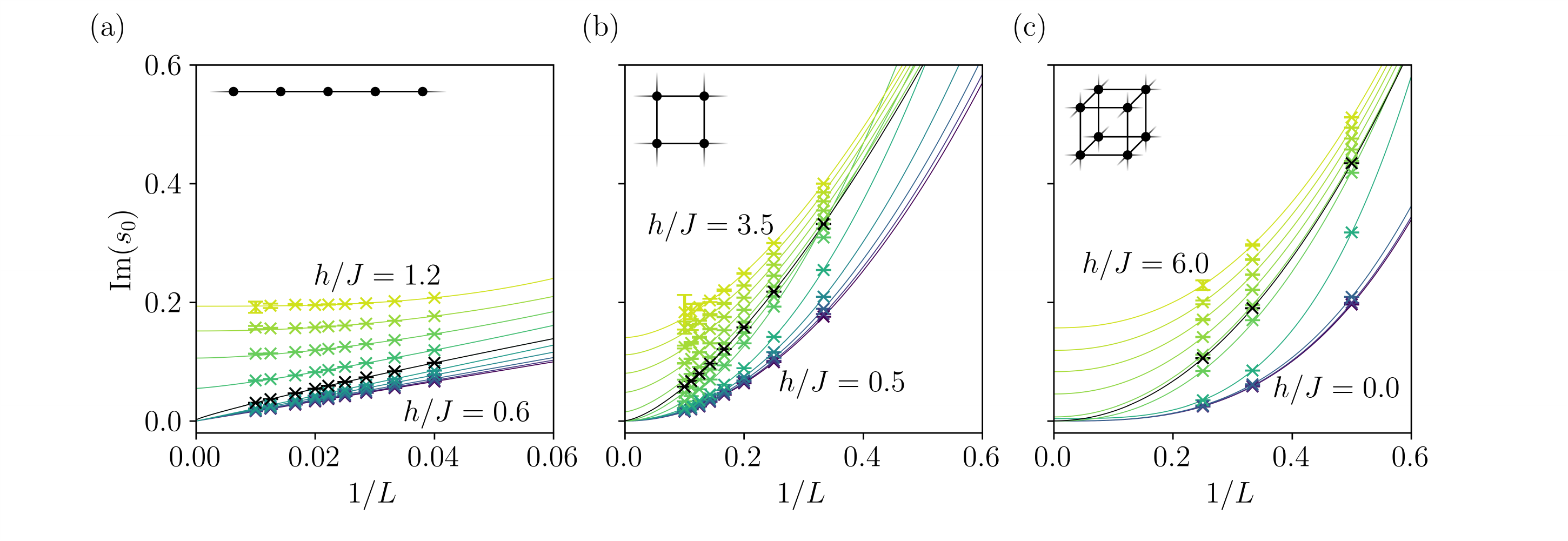}
 \caption{\label{lyzplot} Extraction of zeros from the cumulants of the magnetization. (a) Extracted zeros for a linear Ising chain in different magnetic fields, $h=0.6, 0.7, 0.8, 0.9, 0.95, 1.0, 1.05, 1.1, 1.15, 1.2J$ (starting from the lower curve), as a function of the inverse system size, $1/L$\revision{, with $L=30,40,50,25,35,45,60,80,100$}. The solid lines are the finite-size scaling ansatz in Eq.~(\ref{eq:extrapolationFunction}), which allows us to determine the value in the thermodynamic limit, where $1/L$ approaches zero. \revision{The data points in black correspond to the value of $h/J$, which is closest to the critical field.} (b) Similar results for a two-dimensional square lattice with the following values of the magnetic field, $h=0.5, 1.0, 1.5, 2.0, 2.5, 2.9, 3.0, 3.05, 3.1, 3.2,  3.3, 3.4, 3.5J$ (starting from the lower curve)\revision{ with $L=3,4,5,6,7,8,9,10$}. (c) Results for a cubic lattice in three dimensions with $h=0.0,1.0,2.0,4.0,5.0,5.16,5.2, 5.4, 5.6, 5.8, 6.0J$ (starting from the lower curve)\revision{ with $L=2,3,4$}.}
\end{figure*}

\subsection{Neural network quantum states}

To find the ground state of the system together with the moments and cumulants of the magnetization, we use neural network quantum
states. The neural network quantum states are variational states of the form
\begin{equation}
 \psi_{\vec \theta}(\vec \sigma) = \langle \vec \sigma |\psi_{\vec \theta}\rangle,
\end{equation}
where the vector $\vec \theta$ contains the variational parameters that we need to determine to minimize the energy and thereby find the ground state. The neural network provides a compressed algorithmic representation of the coefficients of the wavefunction, and it takes a spin configuration in the computational basis as the input, and outputs the wave function in response. The energy is minimized using stochastic reconfiguration, which is an approximate imaginary time-evolution within the variational space of the neural network. Neural network state methodologies have been extended to the time-evolution of quantum systems~\cite{Carleo2017, Lin2022scaling, Gutirrez2022realTime}, quantum state tomography~\cite{Torlai2018neuralnetwork, koutny2022neural, neugebauer2020neural}, as well as finite-temperature equilibrium physics~\cite{irikura2020neural, hendry2020neural, nomura2021purifying}.
Importantly, while many other approaches are not able to exploit the computational power of massive parallel computing, neural network quantum states can be implemented with modern graphics processing units.
The energy is evaluated by sampling over the wave function as
\begin{equation}
  \langle \hat \Ham \rangle = \frac{\sum_{\vec \sigma \vec \sigma'} \psi^*(\vec \sigma) \langle \vec \sigma |\hat \Ham |\vec \sigma'\rangle \psi(\vec \sigma') }{ \sum_{\vec \sigma'} |\psi(\vec \sigma')|^2} = \sum_{\vec \sigma} P_\psi (\vec \sigma) \Ham_{\text{loc}}(\vec \sigma),
  \label{eq:aveHam}
\end{equation}
where we have defined the probability
\begin{equation}
 P_\psi(\vec \sigma) = \frac{|\psi(\vec \sigma)|^2}{ \sum_{\vec \sigma'} |\psi(\vec \sigma')|^2}
\end{equation}
and the local spin Hamiltonian
\begin{equation}
 \Ham_{\text{loc}}(\vec \sigma) = \sum_{\vec \sigma'} \langle\vec \sigma |\hat \Ham |\vec \sigma'\rangle \frac{\psi(\vec \sigma')}{\psi(\vec \sigma)}.
 \label{eq:locHam}
\end{equation}
Since Eq.~(\ref{eq:aveHam}) is just an average with respect to a normalized probability distribution, Markov-chain Monte Carlo can be used for evaluating the energy and the gradients~\cite{Carleo2019Netket}. It is worth noting that the spin Hamiltonian in Eq.~(\ref{eq:locHam}) is given by only a few terms in the sum, since only nearest neighbors are coupled. We will also need the expectation value of the total magnetization and its moments, which we express as
\begin{equation}
 \langle \hat M_z^n \rangle = \sum_{\vec \sigma} P_\psi(\vec \sigma) M_z^n(\vec \sigma),
\end{equation}
since $\hat M_z$ is diagonal in the computational basis, such that $M_z^n(\vec \sigma) =  ( \langle \vec \sigma |\hat M_z|\vec \sigma\rangle ) ^n= \langle \vec \sigma |\hat M_z^n|\vec \sigma\rangle$. Additional details of these calculations are provided in
Appendix~\ref{app:details}.

\begin{figure*}[t!]
 \centering
 \includegraphics[width = \textwidth]{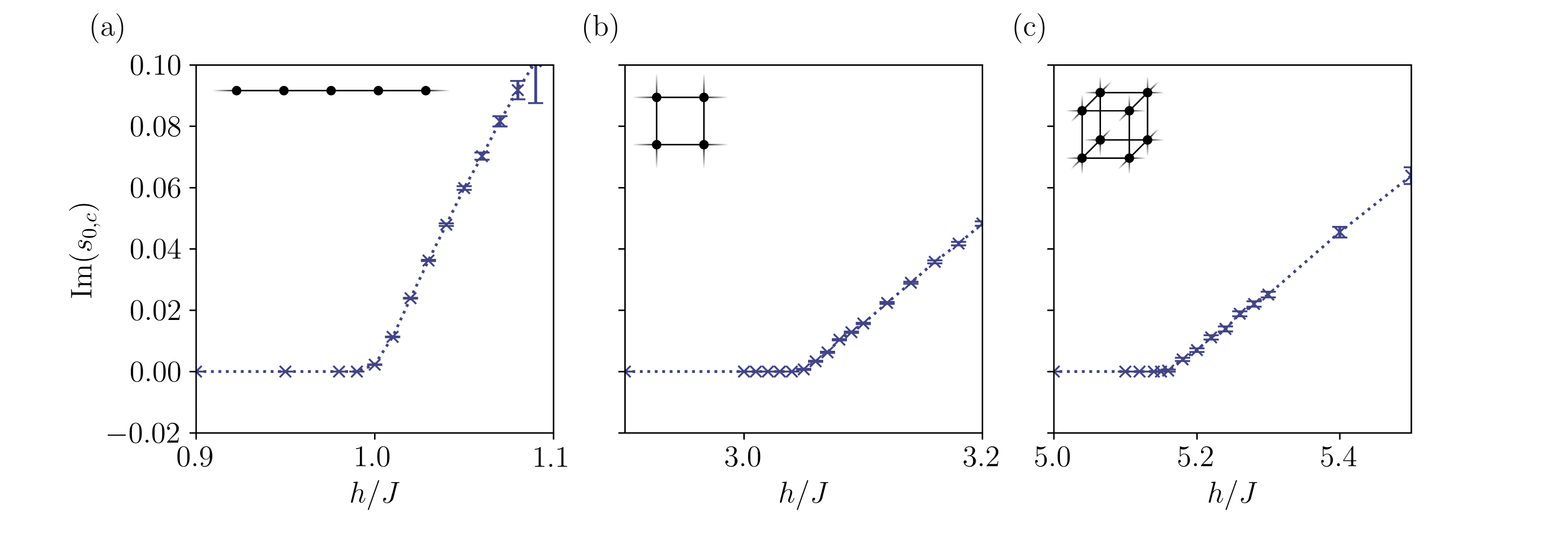}
 \caption{\label{plotOfLee-YangextrapolChainSquareCube} Convergence points of the zeros in the thermodynamic limit. (a) Convergence points for a linear Ising chain as a function of the magnetic field. A quantum phase transition occurs at $h_c=1.00J$, where the curve exhibits a kink, and the zeros reach the real-axis. Above the critical field, the system is in the paramagnetic phase, while it is in the ferromagnetic phase below it. (b,c) Similar results for the two-dimensional square lattice (b) and the cubic lattice in three dimensions (c). 
 }
\end{figure*}

\subsection{Lee-Yang theory}
\label{subsec:leeYang}
The classical Lee-Yang theory of phase transitions considers the zeros of the partition function in the complex plane of the control parameter, for instance magnetic field or inverse temperature~\cite{leeYang1952StatisticalI,leeYang1952StatisticalII,blythe2003theleeyang,bena2005statistical}.
For finite systems, the partition function zeros are situated away from the real axis. However, in case of a phase transition, they will approach the critical value on the real axis in the thermodynamic limit. One may thereby predict the occurrence of a phase transition by investigating the position of the zeros as the system size is increased.  The Lee-Yang theory of phase transitions has found applications in condensed matter physics~\cite{deger2020leeYang, deger2019determination, deger2020leeYang2, deger2018leeYang, kim1998fisher,liu2019fisher}, atomic physics~\cite{mulken2001classification} and particle physics~\cite{barbour1992complex, fodor2002lattice, ejiri2006leeYang, yamamoto2009dense, nagata2015lee, fodor2019trying, wakayama2019lee, dimopoulos2022contribution}.
Recently, it has been extended to the zeros of the moment generating function that describes the fluctuations of the order parameter~\cite{kist2021leeyang,vecsei2022leeyang} and thereby allows for the detection of quantum phase transitions.\revision{ In Ref.~\cite{kist2021leeyang}, the method was implemented for one-dimensional chains, in particular the transverse-field Ising chain and an anisotropic Heisenberg model. In addition, preliminary results for a $J_1$-$J_2$ Heisenberg model in two dimensions were presented. It was  observed that the Lee-Yang approach made it possible to determine the critical points using rather short chains.} Following this approach, we now define  the moment generating function 
\begin{equation}
    \chi(s) = \langle e^{s \hat M_z}\rangle= \frac{1}{g} \sum_{k=1}^g \langle\psi^{(0)}_{k}|e^{s \hat M_z}|\psi^{(0)}_{k}\rangle,
\end{equation}
where $\hat M_z$ is the total magnetization, and $s$ is referred to as the counting field.  Here, we have included the possibility that the system may have $g$ degenerate and normalized ground states that we denote by $|\psi^{(0)}_{k}\rangle$, $k=1,\ldots,g$. Within this framework, the moment generating function plays the role of the partition function in the classical Lee-Yang theory, and the cumulant generating function, $\Theta(s) = \ln \chi(s)$, becomes the corresponding free energy. The moments and cumulants of the magnetization are given by derivatives with respect to the counting field as
\begin{equation}
\langle \hat M_z^n\rangle = \partial_s^n \chi(s)|_{s=0}
\end{equation}
and
\begin{equation}
\langle\!\langle \hat M_z^n\rangle\!\rangle = \partial_s^n \Theta(s)|_{s=0}.
\end{equation}
Importantly, away from a phase transition, the cumulants are expected to grow linearly with the system size, such that the normalized cumulants $\langle\!\langle \hat M_z^n\rangle\!\rangle/N$ converge to finite values as the number of spins $N$ approaches infinity. By contrast, at a phase transition, a different scaling behavior is expected due to as non-analytic behavior of the cumulant generating function at $s=0$ \cite{Karzig2010,deger2018leeYang}. This non-analytic behavior emerges in the thermodynamic limit, if the complex zeros of the moment generating function approach $s=0$. 

To determine the position of the zeros that are closest to $s=0$, we use the cumulant method that was developed in Refs.~\cite{Flindt2013,deger2018leeYang,kist2021leeyang,vecsei2022leeyang,Brandner2017}. In this approach, the zeros of the moment generating function can be determined from the high cumulants of the order parameter. By doing so for different system sizes, we can then find the convergence points in the thermodynamic limit using finite-size scaling~\cite{Flindt2013,deger2018leeYang,kist2021leeyang, vecsei2022leeyang}. The cumulant method allows us to express the zeros in terms of the high cumulants of the magnetization. Moreover, for the transverse-field Ising model, the symmetry, $\hat U^\dag\hat H\hat U= \hat H$, with respect to the unitary operator $\hat U = \prod_i \hat\sigma_i^x$ that flips all spins, implies that all odd cumulants vanish, and in this model the zeros are purely imaginary~\cite{kist2021leeyang, vecsei2022leeyang}. In that case, the zeros that are closest to $s=0$ can be approximated as~\cite{vecsei2022leeyang}
\begin{equation}
 \text{Im}(s_0) \simeq \sqrt{2n (2n + 1) |\llangle \hat M_z^{2n} \rrangle / \llangle \hat M_z^{2n+2} \rrangle|}
 \label{eq:cumulantmeth}
\end{equation}
for large enough cumulant orders, $n\gg1$. Thus, in the following, we find the zeros from the high magnetization cumulants, which we calculate using neural network quantum states, and we ensure that the results from Eq.~(\ref{eq:cumulantmeth}) are unchanged if we increase the cumulant order. We then use the scaling ansatz ~\cite{kist2021leeyang,vecsei2022leeyang}
\begin{equation}
 \text{Im}(s_0) \simeq \text{Im}(s_{0,c}) + \alpha L^{-\gamma}
 \label{eq:extrapolationFunction}
 \end{equation}
to predict the convergence point, $\text{Im}(s_{0,c})$, in the thermodynamic limit, where $L\rightarrow\infty$ is the linear system size. We carry out this procedure for different magnetic fields to find the critical field, where the zeros reach $s=0$, and the system exhibits a phase transition.

\begin{figure*}[t!]
 \centering
 \includegraphics[width = 1.0\textwidth]{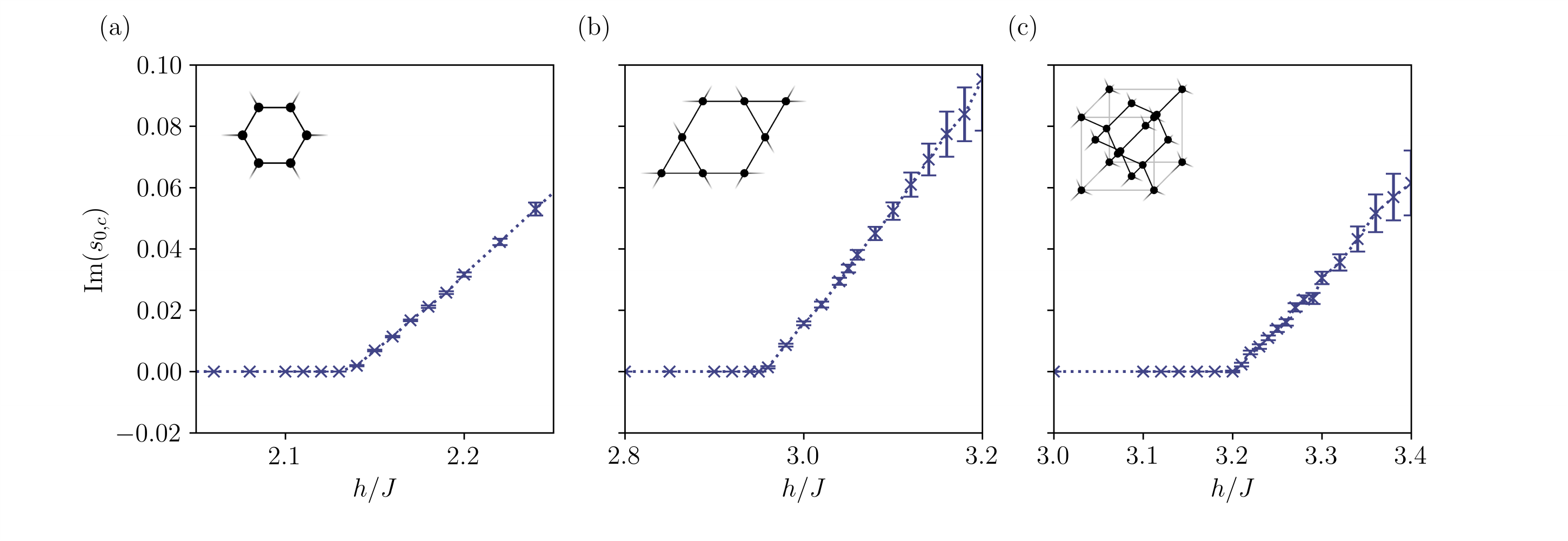}
 \caption{\label{plotOfLee-YangextrapolHoneycombKagomeDiamond} Convergence points of the zeros in the thermodynamic limit. (a) Convergence points for honeycomb lattice as a function of the magnetic field. A quantum phase transition occurs at $h_c\approx 2.14J$, where the curve exhibits a kink, and the zeros reach the real-axis. Above the critical field, the system is in the paramagnetic phase, while it is in the ferromagnetic phase below it. (b,c) Similar results for the Kagome lattice (b) and the diamond lattice (c). }
 \end{figure*}

\section{Results}
\label{sec:results}

\subsection{Extracted zeros}
Figure~\ref{lyzplot} shows zeros obtained for the transverse-field Ising model in one (chain), two (square), and three (cube) dimensions. In each case, we have determined the zeros from Eq.~(\ref{eq:cumulantmeth}) using magnetization cumulants of up to order $n=10$ for a fixed magnetic field and a given system size.  We then obtain the imaginary part of the zeros, and using the finite-size scaling ansatz from Eq.~(\ref{eq:extrapolationFunction}), we find the convergence point in the thermodynamic limit as illustrated in the figure. As an example, we see in Fig.~\ref{lyzplot}a how the zeros eventually reach $s=0$ as we decrease the magnetic field from above to $h\simeq J$, where the system exhibits a quantum phase transition.
In Figs.~\ref{lyzplot}b and~\ref{lyzplot}c, we show similar results for the two-dimensional square lattice and for the three-dimensional cubic lattice. For increased dimensionality, we observe that the quantum phase transitions occurs at higher magnetic fields, as expected for an increasing number of nearest neighbors. In one dimension, we use chains of up to a length of $L=100$. For the two-dimensional square lattices, we consider systems of sizes up to $L\times L= 10 \times 10$, while in three dimensions, the biggest lattice is of size $L\times L\times L =4\times 4 \times 4$. The figure includes small error bars that represent sampling errors in the neural network quantum states. We note that
additional errors could potentially arise from small inaccuracies in the variational ground state.

The results for the three different geometries are combined in Fig.~\ref{plotOfLee-YangextrapolChainSquareCube}, where we show the extracted convergence points as a function of the transverse magnetic field. The extrapolation is performed by a
constrained minimization of $\text{Im}(s_{0,c})$, imposing that the imaginary part is not negative. At large magnetic fields, the systems are in the paramagnetic phase with the spins mostly pointing along the direction of the field. In that case, the zeros of the moment generating function do not converge to $s=0$ in the thermodynamic limit. By contrast, as the magnetic field is lowered, the zeros eventually reach $s=0$, signaling a quantum phase transition. Based on our calculations, we estimate the critical fields to be $h_c = 1.00J$ for the one-dimensional chain, $h_{c} = 3.05J$ for the two-dimensional square lattice, and $h_{c} = 5.16J$ for the three-dimensional cubic lattice. These values are all within less than 1\% difference from other numerical results~\cite{bloete2002cluster}. Below the critical field, the zeros also reach $s=0$, since the system is in the ferromagnetic phase with spontaneous magnetization. In that case, the ground state is two fold-degenerate, and the system will exhibit an abrupt change if a small magnetic field is applied in the $z$-direction.

\subsection{Critical magnetic fields}

We have considered other geometries in two and three dimensions as illustrated in 
Fig.~\ref{plotOfLee-YangextrapolHoneycombKagomeDiamond}, where we show results for a honeycomb lattice, a Kagome lattice, and a diamond lattice.  The honeycomb lattice has two sites per unit cell, and we restrict ourselves to a linear dimension of $L = 8$, which corresponds to $2\times L^2  = 128$ sites. Similarly, for the Kagome lattice, we go up to $L = 6$, while for the diamond lattice, we consider systems of linear size up to $L=4$, which corresponds to $2\times L^3 = 128$ sites. The results in Fig.~\ref{plotOfLee-YangextrapolHoneycombKagomeDiamond} are qualitatively similar to those in Fig.~\ref{plotOfLee-YangextrapolChainSquareCube}, but with different critical fields. In particular, we find  $h_c = 2.14J$ for the honeycomb lattice, $h_c = 2.95J$ for the Kagome lattice, and $h_c = 3.20J$ for the diamond lattice.

The predictions of the critical fields are summarized in Table~\ref{tab:overViewResults}, where we also show results for triangular lattices in two dimensions and face-centred cubic (FCC) and body-centred cubic (BCC) lattices in three dimensions. The results are ordered according to the dimension $D$ as well as the number of nearest neighbors,
the coordination number $C$. In addition, we indicate the maximum linear dimension that we have used, $L_\text{max}$, and the number of sites in a unit cell, $N_\text{cell}$. Those parameters control the maximum number of spins in the lattice that we have considered, $N_\text{max}$. The last column contains the critical magnetic fields that we predict with the combination of Lee-Yang theory and neural network quantum states. We note that our methodology provides accurate predictions even with a rather low number of lattice sites.

\section{Discussion}
\label{sec:discussion}

\subsection{Dimensionality and lattice geometry}

The importance of the lattice geometry and the dimension of the system can be understood from the results in Table~\ref{tab:overViewResults}. The chain and the honeycomb lattice, which have the lowest coordination numbers, also  have the lowest critical fields. The coordination numbers are larger for the Kagome and the square lattices, where each spin has four nearest neighbors, as well as for the triangular lattice with six nearest neighbors, and we see that the critical fields increase accordingly. For the lattices in three dimensions, the coordination numbers and the critical fields are even larger. Despite this general behavior, we also see that lattices with the same dimension and coordination number (the square and Kagome lattices) still have different critical fields, which are directly related to their specific lattice geometries.

\begin{figure*}
 \centering
 \includegraphics[width = 1.0 \textwidth]{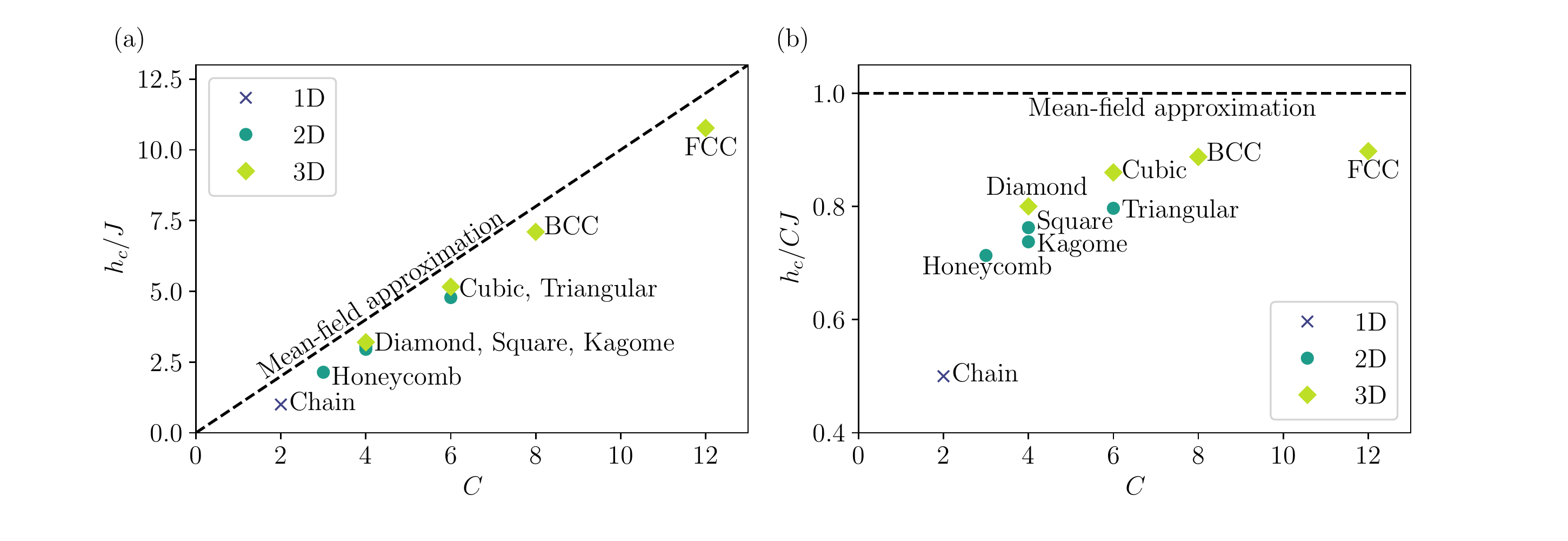}
 \caption{\label{coordnum} Comparison with mean-field theory. (a) The critical fields are shown as functions of the coordination number, $C$. The dashed line is a simple mean-field approximation that directly links the critical field to the coordination number as $h_c^{\text{MF}}=CJ$. (b) The ratio of the critical fields over the mean-field approximation as functions of the  coordination number, $C$. For large coordination numbers and dimensions, the critical fields approach the mean-field approximation indicated with a dashed line.}
\end{figure*}

\subsection{Mean-field approximation}

To better understand the role of the coordination number, we show in Fig.~\ref{coordnum} the critical fields as a function of the coordination number. In Fig.~\ref{coordnum}a, we see the clear trend that the critical fields increase with the coordination number. Indeed, within a simple mean-field approximation, we would expect that the critical field is directly related to the coordination number as $h^{\text{MF}}_c = C J$~\cite{strecka2015brief}. \revision{The physical picture here is that the spins experience a competition between two opposing effects. On the one hand, the external magnetic field tends to align the spins in the $x$-direction. On the other hand, the coupling between them tends to align them along the $z$-axis. The phase transition then occurs when the two effects are equally strong. Since the coupling between neighboring spins is proportional to the coordination number, so is the critical field. Still, this mean-field approximation does not fully capture the actual physics, since it ignores the effects of the lattice geometry. }We show the mean-field approximation with a dashed line in the figure and find good qualitative agreement with our predictions. We also see that our results come closer to the mean-field approximation as the dimension of the system is increased. In particular, it is clear that the critical field for the one-dimensional chain is furthest away
from the mean-field approximation, while the results for the three-dimensional lattices are much closer.
 
To further support these observations, we show in Fig.~\ref{coordnum}b the ratio of the critical fields over the mean-field approximation. This ratio allows us to characterize how the relative deviations from the mean-field prediction decrease for larger coordination numbers. Still, we see that the critical fields are all smaller than the mean-field approximation, which ignores quantum fluctuations. The results for the critical fields in three dimensions are closer to the mean-field approximation as compared with one and two dimensions. This observation is in line with the expectation that mean-field theory becomes more accurate in higher dimensions.

\begin{table}
 \begin{tabular}{ |c | c | c | c | c | c | c | c |}
 \hline
 Lattice & $D$ & $C$ & $L_\text{max}$ & $N_\text{cell}$ & $N_\text{max}$ & $h_c/J$ & Ref.~\cite{bloete2002cluster}\\
 \hline
 Chain & 1 & 2  & 60 & 1 & 60& 1.00 & -\\
 Honeycomb & 2 & 3  & 8 & 2& 128& 2.14 & 2.13\\
 Kagome & 2 & 4  & 6 & 3 & 108& 2.95 & 2.95\\
 Square & 2 & 4  & 10 & 1 & 100& 3.05 & 3.04\\
 Triangular & 2 & 6  & 10 & 1 & 100& 4.78 & 4.77\\
 Diamond & 3 & 4  & 4 & 2 & 128& 3.20 & -\\
 Cubic & 3 & 6  & 4 & 1 & 64& 5.16 & 5.16\\
 BCC & 3 & 8  & 4 & 1 & 64& 7.10 & -\\
 FCC & 3 & 12 & 4 & 1 & 64& 10.8 & -\\
 \hline
 \end{tabular}
 \caption{\label{tab:overViewResults} Summary of critical fields. For each lattice, we indicate the dimension, $D$, and  the coordination number, $C$. We also show the maximum linear dimension, $L_\text{max}$, and the  number of sites per unit cell, $N_\text{cell}$, which give the maximum number of sites that we have used as $N_\text{max} = N_\text{cell} \times L_\text{max}^D$. \revision{The two last columns contain our predictions of the critical field as well as the results from Ref.~\cite{bloete2002cluster}}.}
\end{table}

\section{Conclusions}
\label{sec:conclusion}
We have combined a Lee-Yang theory of quantum phase transitions with neural network quantum states to predict the critical field of the transverse-field Ising model in different dimensions and lattice geometries. Specifically, we have used neural network quantum states to find the ground state of the interacting spin system, which further
makes it possible to extract the cumulants of the magnetization. From these cumulants, we determine the complex zeros of the moment-generating function, which reach the real-axis in the thermodynamic limit if the system exhibits a phase transition. Our method works with rather small systems, which in turn allows us to treat lattices in two and three dimensions. Our predictions agree well with results that were obtained using large-scale quantum many-body methods.
We have also analyzed the differences between our predictions and a simple mean-field approximation, which becomes increasingly accurate for higher coordination numbers and dimensions. 
Thanks to the flexibility of neural network quantum states, the method can potentially treat frustrated problems,
in stark contrast to quantum Monte Carlo approaches that suffer from sign-problems. 
Our results show that the combination of Lee-Yang theories of  phase transitions with neural network quantum states provides a viable way forward to predict the phase behavior of complex quantum many-body systems such as Heisenberg models and fermionic Hubbard models.\revision{ The application of neural network quantum states to fermionic models is currently being developed, which in the future may provide a better understanding of interacting fermionic systems using Lee-Yang theory.}

\acknowledgements

We acknowledge the computational resources provided by the Aalto Science-IT project and the support from the Finnish National Agency for Education (Opetushallitus), the Academy of Finland through grants (Grants No.~331342 and No.~336243) and the Finnish Centre of Excellence in Quantum Technology (Projects No.~312057 and No.~312299), 
and from the Jane and Aatos Erkko Foundation.

\appendix

\section{Details of calculations}
\label{app:details}

All of our calculations were implemented in Netket 3.3~\cite{Carleo2019Netket, Vicentini2022netket}. In one dimension, we found that a restricted Boltzmann machine works well, while in two dimensions, a group convolutional neural network functions better. In three dimensions, we used a simple and shallow symmetric architecture with real weights, which is sufficient, since the transverse-field Ising model is stoquastic. 

In one dimension, we used a simple real restricted Boltzmann machine with a number of hidden units per visible unit of $\alpha = 20$. For each training iteration, 8192 samples were used, taken from 128 parallel chains. The network was trained for 3000 iterations with a learning rate of 0.02, and then for further 1000 iterations with a learning rate of 0.01. Stochastic reconfiguration with a diagonal shift of 0.01 was used.

In two dimensions, we used a group convolutional neural network~\cite{roth2021group, pmlr-v48-cohenc16} defined over the group of all translations with four layers of feature dimension 8 each and complex parameters. We used 32 parallel Markov chains constructed using a Metropolis algorithm with local updates, and we took 1024 samples per iteration step. Stochastic reconfiguration with a diagonal shift of 0.01 was used, and the network was trained with a learning rate of 0.01 for 2000 iterations. If necessary, we trained the network multiple times and chose the network with the lowest variance of the energy.

In three dimensions, we applied a dense symmetric layer with real weights and 40 features to the input, and we then activated it with the ReLu function, which was then summed over to obtain the wave function. We used a local Metropolis update Markov-chain with 128 parallel chains and 8192 samples per training step. A learning rate of 0.002 and stochastic reconfiguration with a diagonal shift of 0.01 were applied. We then trained the network for 2000 iterations. If necessary, we ran this training multiple times for the same configuration (system size and magnetic field), and we chose the network parameters that resulted in the lowest variance of the energy, so that the network was as similar as possible to a \revision{ground state} of the Hamiltonian.

We evaluated the moments of the magnetization using regular sampling with an unbiased Markov chain, since
\begin{equation}
 \langle \hat M_z^n \rangle = \sum_{\vec \sigma} P_\psi(\vec \sigma) M_z^n = \sum_{\vec \sigma} P_\sigma(\vec \sigma)\left[ \sum_i \sigma_i \right]^n.
\end{equation}
For the two- and three-dimensional lattices with up to $N_\text{max} = 128$ sites, we took $100 \times 1024 \times 128 \simeq 13 \text{ million}$ samples. For the one-dimensional lattice, we took up to $1000 \times 1024 \times 128 \simeq 270 \text{ million}$ samples. For the sampling, we used 128 parallel chains and discarded the first 64 entries. From the  moments, we then obtained the cumulants using a standard recursion relation between them.

\revision{
\section{Error analysis}
\begin{figure}
 \centering
 \includegraphics[width = 0.48\textwidth]{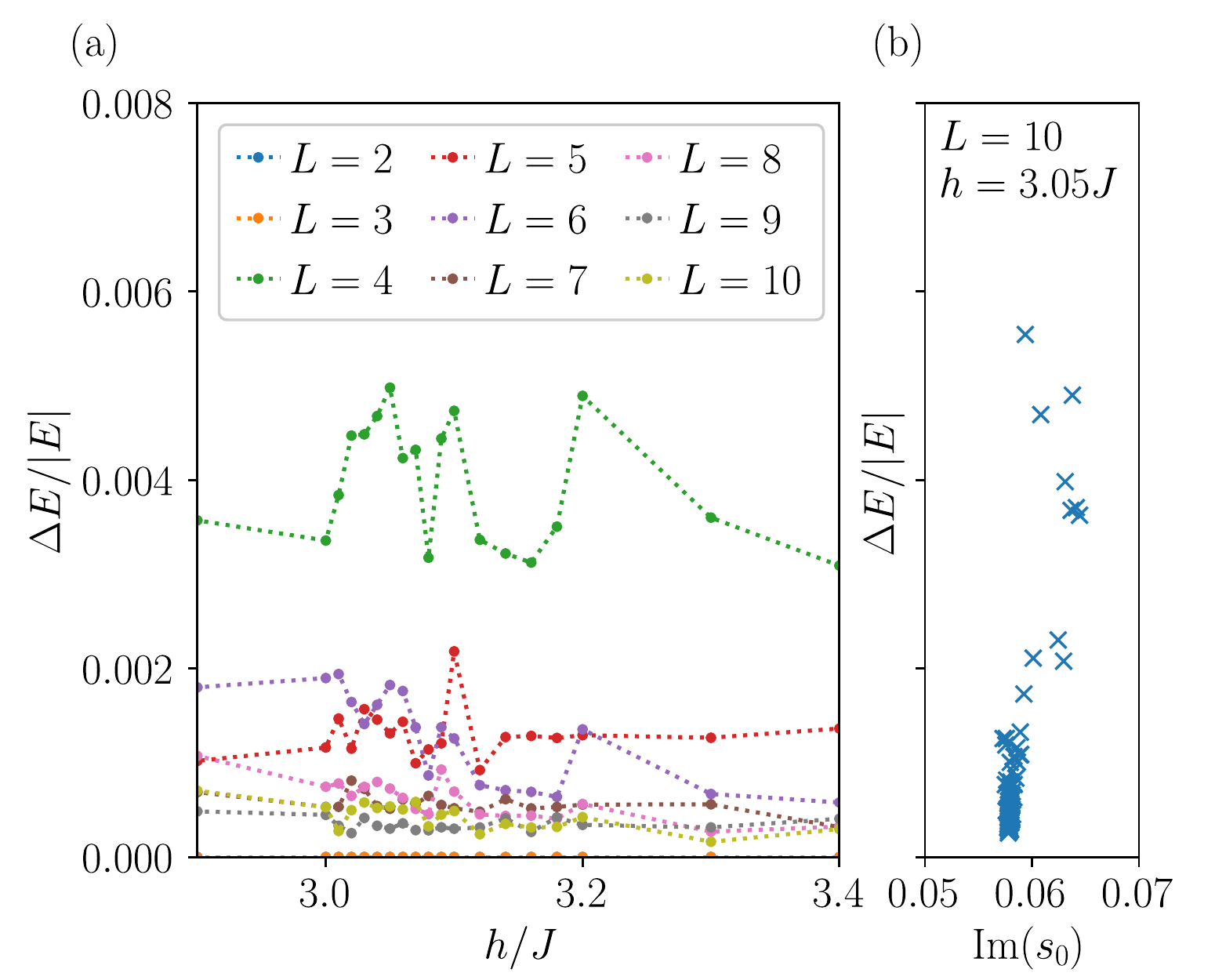}
 \caption{\label{fig:erroranalysis}\revision{ Error analysis. (a) Relative error in the ground state energy for the square lattice of different sizes as a function of the field strength. (b) Zeros obtained for 100 neural network quantum states with $L=10$ and $h/J = 3.05$ as a function of the error. For $\Delta E / |E|   < 0.001$, the zeros have converged.}}
\end{figure}

The position of the zeros can be slightly imprecise for several reasons. For example, the ground state is never completely accurate, which may lead to errors in the zeros. The Monte Carlo sampling itself may also lead to inaccuracies. In addition, one  has to ensure that only the closest zeros contribute to the cumulants by using a sufficiently high cumulant order. The errors from the Monte Carlo sampling are statistical in nature and can easily be quantified. Regarding the ground state, we make sure that it has converged so that the zeros obtained with the cumulant method remained unchanged. In Fig.~\ref{fig:erroranalysis}(a) we show the relative error in the ground state energy for the square lattice, and we see that they are small for all field strengths. In Fig.~\ref{fig:erroranalysis}(b), we show the extracted zero for different relative errors, and we see how the position of the zero converges as the error is reduced. These results were obtained by running the algorithm for finding the ground state ten times, taking ten snapshots each time at different stages of the training. These 100 data points have different errors in the energy as shown in Fig.~\ref{fig:erroranalysis}(b) together with the extracted zero. As the error is reduced, we see a clear convergence of the zero. In particular, for $\Delta E/|\langle E \rangle | < 0.001$, the position of the zero remains the same as the error is further reduced. 
}
\FloatBarrier
\bibliography{Ref-Lib}

\end{document}